\documentstyle[12pt,epsf]{article}
\begin{document}
\hfill \vbox{\hbox{FTUV/95-11} \hbox{IFIC/95-11} \hbox{TK 95 06}}

\vspace{1cm}

\begin{center}
{\large
{\bf CHIRAL SYMMETRY CONSTRAINTS ON THE

\smallskip

$K^+$ INTERACTION WITH THE NUCLEAR PION CLOUD}}

\end{center}

\vspace{1.5cm}
\begin{center}
{\large{Ulf-G. Mei{\ss}ner$^{\ast}$\footnote{electronic address:
meissner@pythia.itkp.uni-bonn.de},
E. Oset$^{\ast \ast}$\footnote{electronic address: oset@evalvx.ific.uv.es}
and
A. Pich$^{\ast \ast}$\footnote{electronic address: pich@papageno.ific.uv.es}}}
\end{center}

\vspace{1.0cm}

\begin{center}

{\small{\it $^{\ast}$Institut f\"ur Theoretische Kernphysik, Universit\"at
Bonn,\\
Nussalle 14-16, D-53115 Bonn, Germany}}

\smallskip

{\small{\it $^{\ast \ast}$Departament de F\'{\i}sica Te\`orica and IFIC, 
Universitat de Val\`encia--CSIC\\
Dr. Moliner 50, E-46100 Burjassot (Val\`encia), Spain}}

\end{center}

\vspace{2cm}

\begin{abstract}
{\small{The real part of the $K^{+}$ selfenergy for the interaction of the
$K^{+}$ with the pion nuclear cloud is evaluated in lowest order of chiral
perturbation theory and is
found to be exactly zero in symmetric nuclear matter.
This removes uncertainties in that quantity found in former
phenomenological analyses
and is supported by present experimental data on $K^{+}$ nucleus scattering.}}
\end{abstract}

\vspace{4cm}

\noindent           
March 1995

\newpage

Systematic discrepancies between the impulse approximation (IA) and the data
in $K^{+}$ nucleus scattering \cite{1,2} have stimulated work searching for
the contribution of the $K^{+}$ interaction with the nuclear pion cloud
\cite{3,4,5} among others \cite{6}. The relatively weak strength of the
$K^{+}N$ interaction led to the suggestion in \cite{3} that the $K^{+}$
interaction with the virtual pion cloud in the nucleus could provide
a correction of the size needed to explain the discrepancies. The idea was
elaborated further in \cite{4} and the contribution to the $K^{+}$ nucleus
selfenergy, $\delta  \Pi$, from the virtual pions was evaluated. Further work
was done in \cite{5} improving the work of \cite{4} in two aspects:
i) the static approximation of \cite{4} was improved by using a covariant
formalism and ii) further contributions to the imaginary part of the $K^{+}$
selfenergy were found which increased appreciably this magnitude.
It was also found in \cite{5}
that the imaginary part of the $K^{+}$ selfenergy
alone provided sufficient strength to account for the discrepancies of the IA
with the data, account taken of extra nuclear corrections stemming from
correlations and evaluated in \cite{2}. However, both in \cite{4} and \cite{5},
which used the same $K^{+} \pi$ amplitude, it was shown that the real part of
the $K^{+}$ selfenergy was not precisely determined and the results depended
strongly on the off--shell extrapolation of the $K^{+} \pi$ amplitude. These
uncertainties in $\delta \Pi$ resulted in large uncertainties in the $K^{+}$
nucleus cross section at small $K^{+}$ energies, while at large energies
$(T_{K^{+}} \sim 500-600$ MeV) these uncertainties were drastically reduced
(see fig. 1). On the other hand, the imaginary part of the $K^{+}$ selfenergy
produced sizeable corrections in the large energy region while, for obvious
reasons of phase space, it provided no correction at small energies. Comparison
with the data at low energies (see fig. 1) led to the conclusion in \cite{5}
that the $K^{+}$ nucleus data were best reproduced with the IA selfenergy and
a contribution from the pion cloud such that $\mbox{\rm Re}(\delta\Pi) \simeq
0$.

In the present work we show that in the evaluation of
$\mbox{\rm Re}(\delta\Pi)$ there are
chiral terms, additional to the terms from the $K^{+}\pi$ interaction used in
\cite{4,5}, which cancel exactly these latter terms leading to
$\mbox{\rm Re}(\delta\Pi)=0$.

The $K^{+}$ selfenergy in \cite{5}, for the interaction of a $K^{+}$ with
symmetric matter is given by

\begin{equation}
\delta \Pi (k) = i \int \frac{d^{4}q}{(2 \pi )^{4}} \,\delta D (q)
\,\frac{3}{2} t^{0} (k,q; k,q) \, ,
\end{equation}

\noindent
with

\begin{equation}
\delta D(q) = D (q) - D_{0} (q) - \rho
\left(\frac{\partial D(q)}{\partial \rho}\right)_{\rho = 0}
\end{equation}

\noindent
where $k$ is the $K^{+}$ momentum, $t^{0}$ the isoscalar $K^{+}\pi$ amplitude
and $D_{0} (q), D (q)$ the free pion propagator and the pion propagator in the
nuclear medium, respectively.
Note that in eq. (1) one evaluates the interaction
of a $K^{+}$ with a pion in the medium and makes two subtractions: the one of
the free pion, because it is already contained in the free $K^{+}$
mass,  and the
terms linear in $\, \, \rho$, the nuclear density, because they are contained
in the IA $K^{+}$ selfenergy, $\Pi = t^{0}_{K N} \; \rho$. One thus picks up
corrections of type $\rho^{\alpha} \, (\alpha > 1)$, essentially $\rho^{2}$
terms in eq. (1). They are depicted in figs. 2 a, d.

The novelty introduced in this paper is the realization that in a systematic
chiral perturbation expansion there are terms like those shown in
figs. 2b, c, e, f which appear at the same order of
the terms in figs. 2a, 2d and which cancel
them exactly. This fact was already noticed in \cite{8}, studying the
interaction of pions with the nuclear pion cloud, where a
{\it partial} cancellation
of the analogous terms to fig. 2, substituting the $K^{+}$ by a pion, was
found.
A different technical approach to that problem with a different interpretation
of the results of \cite{8} is given in \cite{9}.

In addition to the diagrams of fig. 2 one has the corresponding ones to a, b,
c substituting one or the two $ph$ excitations by a $\Delta h$ excitation.
However, there are no corrections of the type of diagrams d, e, f with $\Delta
h$ excitation since there is no Fermi sea of $\Delta 's$. To clarify this point
let us see the meaning of the arrows in the baryon lines. In the diagrams
2 a, b, c the meaning of the arrows is the standard one for particles and
holes.
In figs. 2d, e, f, one of the
lines pointing down correspond to a hole line while the other one corresponds
to the Pauli correction to the nucleon propagator, $2\pi i n (\vec{p}) \delta
(p^{0} - \epsilon (\vec{p}))$, which appears when the terms linear in $\rho$
are subtracted in the equivalent diagrams with the bubble representing a
standard $ph$ excitation (see appendix of ref. \cite{7} for the appropriate
many body details). In the absence of a Fermi sea of $\Delta$ 's,
one has $\; n_{\Delta}
(\vec{p})=0$ and the terms d, e, f with $\Delta 's$ do not appear.
We will omit other many-body details here since they will be
unnecessary to prove that $\mbox{\rm Re}(\delta\Pi)=0$.

In order to evaluate the diagrams of fig. 2 we need the $KK\pi \pi$ vertex,
the $\pi \bar{N}N$ vertex and the $\bar{N}NKK\pi$ contact term.
All of them are directly obtained from the standard Chiral Perturbation Theory
Lagrangian \cite{10,11,12,13}, which contains the most general low-energy
interactions of the pseudoscalar and baryon octets, consistent with the chiral
symmetry properties of QCD. At lowest order in derivatives and quark masses it
reads:

\begin{equation}
{\cal L}_{ChPT} = {\cal L}_{2} + {\cal L}_{1}^{(B)} + \cdots
\end{equation}

\begin{equation}
{\cal L}_{2}= \frac{f^{2}}{4} \langle\partial_{\mu} U^{+} \partial^{\mu} U +
M(U+U^{+})\rangle \, ,
\end{equation}

\begin{eqnarray}
{\cal L}_{1}^{(B)} &=&\langle \bar{B} i \gamma^{\mu} \nabla_{\mu} B \rangle
- M_{B} \langle \bar{B} B\rangle \nonumber\\
& +&
\frac{D + F}{2} \langle \bar{B} \gamma^{\mu} \gamma_{5} u_{\mu} B\rangle +
\frac{D -F}{2} \langle\bar{B} \gamma^{\mu} \gamma_{5} B u_{\mu}\rangle\, ,
\end{eqnarray}

\noindent
where the symbol $\langle \; \rangle$
denotes the flavour trace of the $SU(3)$ matrices
and the dots in eq. (3) stand for interactions with more than two baryons and
higher-order terms in the momentum expansion.
The $3 \times 3$ unitary matrix
\begin{equation}
U (\phi ) = u (\phi )^{2} = \exp (i \sqrt{2} \Phi /f)
\end{equation}
\noindent
gives a very convenient parametrization of the pseudoscalar Goldstone fields
\begin{equation}
\Phi (x) \equiv \frac{\vec{\lambda}}{\sqrt{2}} \vec{\phi}=
\left(
\begin{array}{ccc}
\frac{1}{\sqrt{2}} \pi^{0} + \frac{1}{\sqrt{6}} \eta_{8} & \pi^{+} & K^{+}\\
\pi^{-} & - \frac{1}{\sqrt{2}} \pi^{0} + \frac{1}{\sqrt{6}} \eta_{8} & K^{0}\\
K^{-} & \bar{K}^{0} & - \frac{2}{\sqrt{6}} \eta_{8}
\end{array}
\right) \, .
\end{equation}
The baryon octet is collected  in the $3 \times 3$ matrix $B$ \cite{11,12,13},
which in our particular case, where only protons and neutrons are involved,
simplifies to
\begin{equation}
B(x)=
\left(
\begin{array}{lll}
0 & 0 & p\\
0 & 0 & n\\
0 & 0 & 0
\end{array}
\right)
= \frac{1}{2}(\lambda_4 + i \lambda_5) \, p +
\frac{1}{2}(\lambda_6 + i \lambda_7) \, n \, .
\end{equation}
The matrix $M$ contains the explicit breaking of chiral symmetry generated by
the non-zero quark masses. In the isospin limit $(m_{u} = m_{d})$, it takes the
simple form
\begin{equation}
M=
\left(
\begin{array}{ccc}
m^{2}_{\pi} & 0 & 0\\
0 & m_{\pi}^{2} & 0\\
0 & 0 & 2 m_{K}^{2}-m_{\pi}^{2}
\end{array}
\right) \, .
\end{equation}
The baryon covariant derivative $\nabla_{\mu}$ and $u_{\mu}$ are given by
\begin{equation}
\nabla_{\mu} B = \partial_{\mu} B + [ \Gamma_{\mu} , B] \; ; \; \qquad
\Gamma_{\mu} = \frac{1}{2} (u^{+} \partial_{\mu} u + u \partial_{\mu} u^{+})
\; ;
\end{equation}
\begin{equation}
u_{\mu} = i u^{+} \partial_{\mu} U u^{+} \; .
\end{equation}
At lowest order, $f$ equals the pion decay constant, $f = f_{\pi} = 92.4 \;$
MeV, while $D$ and $F$ are the usual parameters of semileptonic hyperon decays,
which satisfy $D + F = g_{A} = 1.257 \pm 0.003$.

Expanding $u(\phi )$ in a power series in $\Phi$, we easily obtain the needed
interactions:

\begin{equation}
{\cal L}_{2} = \frac{1}{2} \langle\partial_{\mu} \Phi \partial^{\mu} \Phi - M
\Phi^{2}\rangle
+ \frac{1}{12 f^{2}} \langle (\partial_{\mu} \Phi \Phi - \Phi \partial_{\mu}
\Phi
)^{2} + M \Phi^{4} \rangle + O (\Phi^{6}) \, ,
\end{equation}

\begin{equation}
u_{\mu} = - \frac{\sqrt{2}}{f} \partial_{\mu} \Phi + \frac{\sqrt{2}}{12 f^{3}}
(\partial_{\mu} \Phi \Phi^{2} - 2 \Phi \partial_{\mu} \Phi \Phi + \Phi^{2}
\partial_{\mu} \Phi ) + O (\Phi^{5}) \, .
\end{equation}
Only the $O(\Phi^{4})$ interaction terms in ${\cal L}_{2}$ and the $(D + F)$
and $(D - F)$ couplings in ${\cal L}_{1}^{(B)}$ [i.e., the $O (\Phi )$ and
$O ( \Phi^{3})$ terms in $u_{\mu}$] contribute to the processes in fig. 2. The
relevant baryon vertices are

\begin{eqnarray}
{\cal L}_{1}^{(B)} &=& \frac{D + F}{2} \biggl\{
\bar{p} \gamma^{\mu} \gamma_{5}
u_{\mu}^{11} p + \bar{n} \gamma^{\mu} \gamma_{5} u_{\mu}^{22} n
+ \bar{n} \gamma^{\mu} \gamma_{5} u_{\mu}^{21} p +
\bar{p} \gamma^{\mu}
\gamma_{5} u_{\mu}^{12} n \biggr\} \nonumber\\
&+& \frac{D - F}{2} \biggl\{
\bar{p} \gamma^{\mu} \gamma_{5} u_{\mu}^{33} p +
\bar{n} \gamma^{\mu} \gamma_{5} u_{\mu}^{33} n \biggr\} \, ,
\end{eqnarray}
where $u_{\mu}^{ij}$ denote the $(i, j)$ element of the matrix $u_{\mu}$.

The terms in fig. 2 can be calculated in a systematic way from the Lagrangians
of eqs. (12-14) and the cancellations appear in a subtle way in
spin-isospin saturated nuclear matter. In order to show the cancellations it
is useful to introduce the
effective vertex of fig. 3 which allows then to rewrite the diagrams of the
first row of fig. 2 as in fig. 4. We are only interested in the kinematics of
the present case where $k^{\mu} = k'^{\mu} \; , \; k^{2}= k'^{2} = m_{K}^{2}$,
and the results in the following refer to this particular case.

In the effective vertex for a $\pi^{0}$ the
$pp$ and $nn$ matrix elements can be cast into an isospin form given by
$[N^T \equiv (p \, \, n)]$

\begin{equation}
-it_{0}\equiv \frac{1}{12 f^{3}} \bar{N} \gamma^{\mu} \gamma_{5} q_{\mu}
\, \biggl\{ - (D + F) \frac{1}{2} \tau_{3} +
 (D + F) \frac{1}{2} (1 + \tau_{3})-
(D - F) \biggr\} N \, ,
\end{equation}
where the first term in the bracket corresponds to the one half of the pion
pole term in the effective vertex and the other two to the contact term. The
terms with $\tau_{3}$ cancel in eq. (15) and one obtains only the isoscalar
contribution

\begin{equation}
-it_{0} = \frac{3 F- D}{2}\,
\frac{1}{12 f^{3}} \,\bar{N} \gamma^{\mu} \gamma_{5} N q_{\mu} \, .
\end{equation}
It is then trivial to see the cancellation in fig. 4 for the neutral pions
since the $\pi^{0} \bar NN$ vertex below the effective
vertex in the $ph$ excitation
has a $+, -$ sign for $p$ or $n$, and the sum over protons and neutrons
cancels in symmetric nuclear matter.

For the case of charged pions it is easy to see that the sum of the matrix
elements of the effective vertex for a $\pi^{+}$ and a $\pi^{-}$ vanishes
identically. In this case the effective vertices are

\begin{equation}
-it_{-} = \frac{1}{12\sqrt{2} f^{3}}
\bar{N} \gamma^{\mu} \gamma_{5} \tau_{-} N \, \biggl\{ - (D + F)
 \frac{q_{\mu}}{q^{2}- m_{\pi}^{2}} (q^{2} - m_{\pi}^{2} -
6 q \cdot k) + (D + F) (q_{\mu} - 3k_{\mu}) \biggr\}
\end{equation}

\begin{equation}
-it_{+} = \frac{1}{12\sqrt{2} f^{3}}
\bar{N} \gamma^{\mu} \gamma_{5} \tau_{+} N
\, \biggl\{ - (D + F)
\frac{q_{\mu}}{q^{2} - m_{\pi}^{2}} (q^{2} - m_{\pi}^{2} +
6q \cdot k) + (D + F) (q_{\mu} + 3 k_{\mu})\biggr\}
\end{equation}

\noindent
with $q^{\mu}, k^{\mu}$ the pion and kaon momenta, respectively.
Again the first term in the bracket corresponds to  one half of the
pion pole term and the second one to the contact term. Since the
$\pi \bar NN$ vertex
below the effective vertex has the same sign in $p, n$ or $n, p$ excitation
($\sqrt{2}$ isospin factor for charged pions), then the cancellation in fig.
4 holds exactly provided we have equal number of protons and neutrons in the
system.

The isospin formalism used is particularly useful. The cancellation of the
diagrams in figs. $2d, e, f$ follows the same arguments as before. On the
other hand it is trivial to show the cancellation of the $\Delta$ terms. For
this purpose we define the effective vertex at the quark level with quarks
$u, d$. These vertices in all the isospin combinations have the same expression
as in eqs. (16-18) up to some normalization factor. Then the sum of matrix
elements of this effective vertex for $\pi^{+}$ and $\pi^{-}$ vanishes
identically while for a $\pi^{0}$ it is an isoscalar. If we take $SU(4)$ wave
functions for $N$ and $\Delta$ then we recover the former matrix elements for
the
nucleons while the effective $\bar N \Delta KK \pi^{0}$ vertex is identically
zero
and the sum of these effective vertices with $\pi^{+}$ and $\pi^{-}$ also
vanishes. Since the lower vertex $\pi \bar N \Delta$ has the same isospin
matrix
element for all charged pion cases ($\sqrt{1/3}$ isospin coefficient), the
cancellation for charged pions holds also exactly in symmetric nuclear matter.
We have also checked these cancellations
using the effective chiral Lagrangian
incorporating explicitly the baryon decuplet \cite{11,15}.

In summary, the approach followed here is different and more systematic than
the phenomenological one followed in refs. \cite{4,5}. While in this latter
work there were uncertainties in $\mbox{\rm Re}(\delta\Pi)$
tied to the off--shell
extrapolation of the $K\pi$ amplitude, we have shown here that in lowest
order in chiral perturbation theory there is an exact cancellation between
different terms, for symmetric nuclear matter,
 and $\mbox{\rm Re}(\delta\Pi)$ is zero.
Furthermore, the empiral analysis carried
out in ref. \cite{5} clearly favoured
$\mbox{\rm Re}(\delta\Pi)=0$. This is certainly
a welcome feature. Obviously one may
wonder what would happen if one goes to higher orders in the chiral expansion.
The $K^{+}$ sector might be a privileged one for convergence of the chiral
expansion, since one is far away
from resonances in the $K^{+} \pi$  channel (the $K^{\ast}(892)$ considered in
\cite{4} has a negligible influence in the $K\pi$ amplitude in the range
of energies considered in \cite{4,5})
and there are no resonances in the $K^{+}N$
channel. In spite of that, a systematic expansion at higher orders,
(as done in ref. \cite{16} for the elastic $KN$ scattering matrix) looking
also for $\mbox{\rm Im}(\delta\Pi)$ as an alternative to the empirical
evaluation
done in refs. \cite{4,5}, would be an interesting step forward.

\vspace{1.5cm}

\noindent {\large {\bf Acknowledgements:}}

\medskip

One of us (E.O.) would like to acknowledge useful discussions with
C. Garc\'{\i}a--Recio, J. Nieves and W. Weise.
U.G.M. wishes to acknowledge the
hospitality of the University of Valencia where part of the work was done.
This work has been partially supported by CICYT contract numbers
AEN 93-1205 and AEN 93-0234.

\bigskip

\bigskip

\noindent{\large{\bf Figure Captions:}}

\bigskip

\begin{enumerate}

\item[Fig. 1:]
 Comparison with the data for the total cross section of the impulse
approximation (IA) and the (IA) plus the $K^{+}$ selfenergy
form the interaction
with the pion cloud, $\delta \Pi$. The results are from ref. \cite{5} and the
data form ref. \cite{14}. The curve labelled IA shows the IA results. The
curves labelled I, III, IV correspond to different values of
$\mbox{\rm Re}(\delta\Pi)$
coming from different $K^{+} \pi$ off--shell extrapolations and the same
$\mbox{\rm Im}(\delta\Pi)$ \cite{5}. The curve labelled IV
(solid line) corresponds to
$\mbox{\rm Re}(\delta\Pi)=0$.

\medskip

\item[Fig. 2:]
 Diagrams which enter the evaluation of eq. (1) (a, d) plus the extra
terms (b, c, e, f) which appear in the chiral expansion at lowest
order. Solid and dashed lines denote nucleons and mesons, respectively.
In-coming and out-going mesons are kaons, whereas the exchanged mesons
are pions.

\medskip

\item[Fig. 3:]
 Diagrammatic definition of the effective $\bar{N}NKK \pi$ vertex,
 depicted by the circle-cross.

\medskip

\item[Fig. 4:] Detail of the diagrams
of figs. 2a, b, c for the different isospin
combinations in terms of the effective vertex,
together with their cancellation.

\end{enumerate}

\newpage
\hbox{}\vspace{5cm}
\begin{figure}[bht]
\centerline{
\epsfxsize=6in
\epsffile{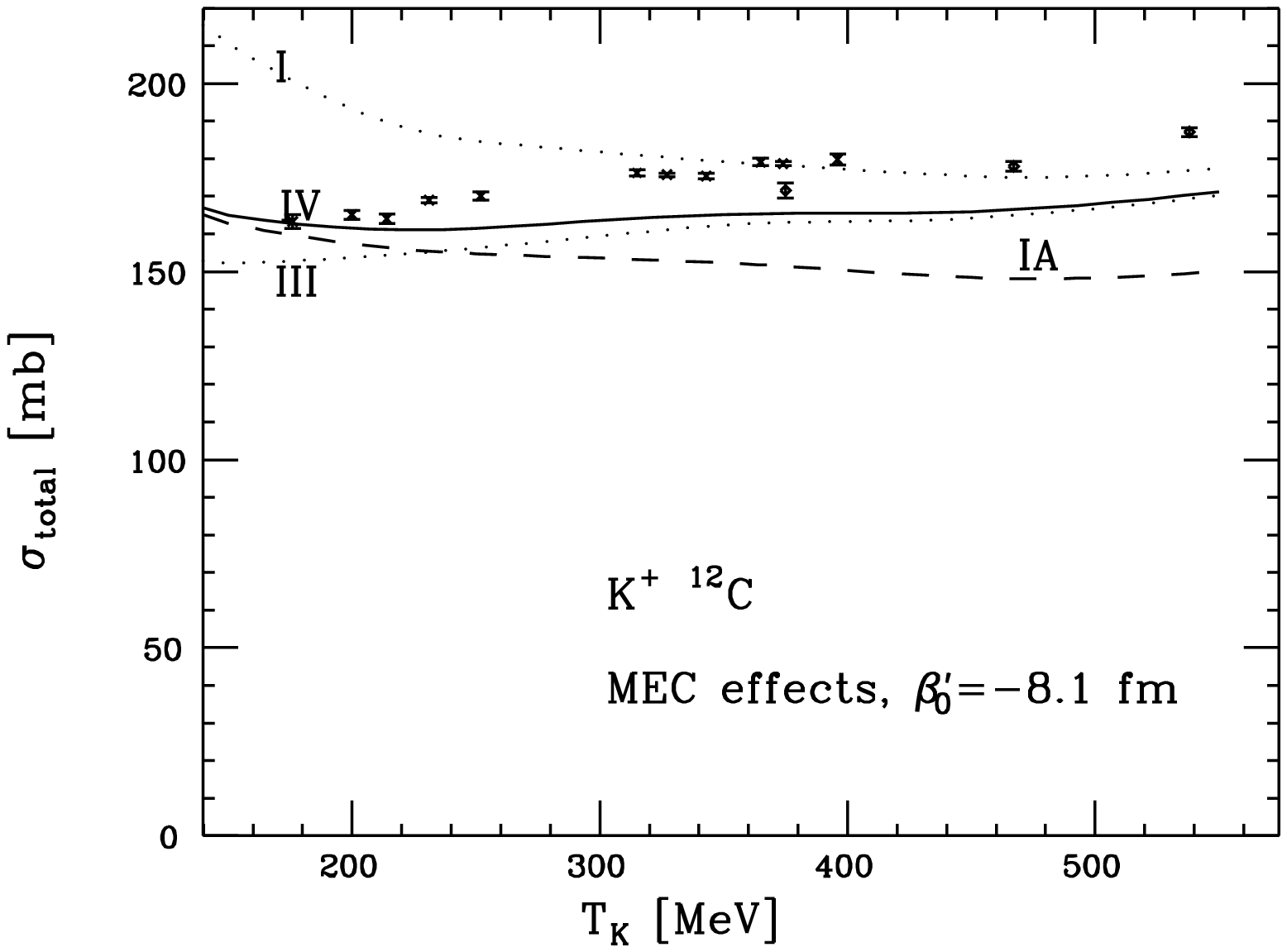}
}
\bigskip\bigskip\bigskip\bigskip\bigskip\bigskip
\bigskip\bigskip\bigskip\bigskip\bigskip\bigskip\bigskip
\centerline{\Large Figure 1}
\end{figure}


\begin{figure}[bht]
\centerline{
\epsfxsize=6in
\epsffile{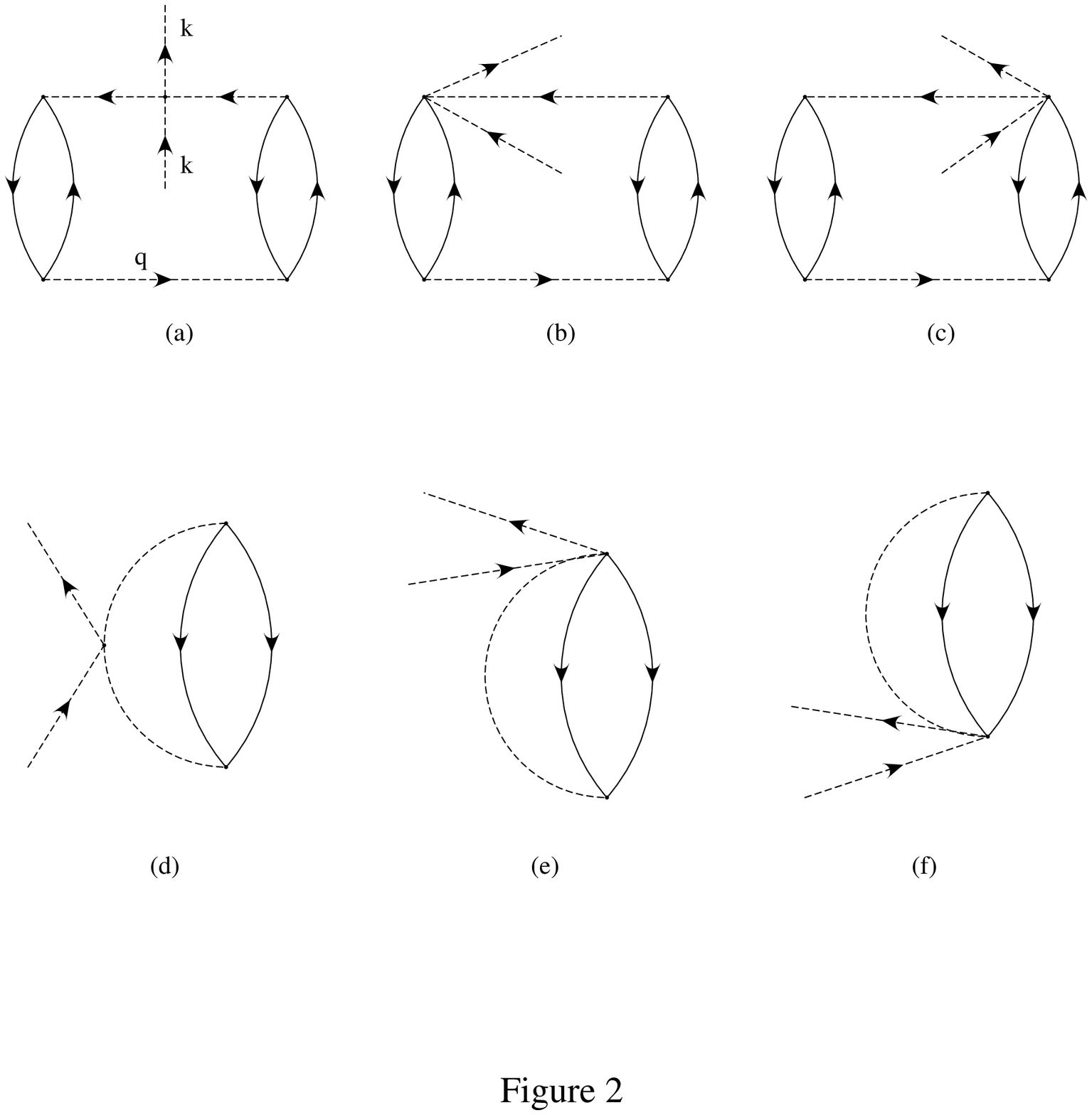}
}
\end{figure}


\newpage
\begin{figure}[bht]
\centerline{
\epsfxsize=12cm
\epsffile{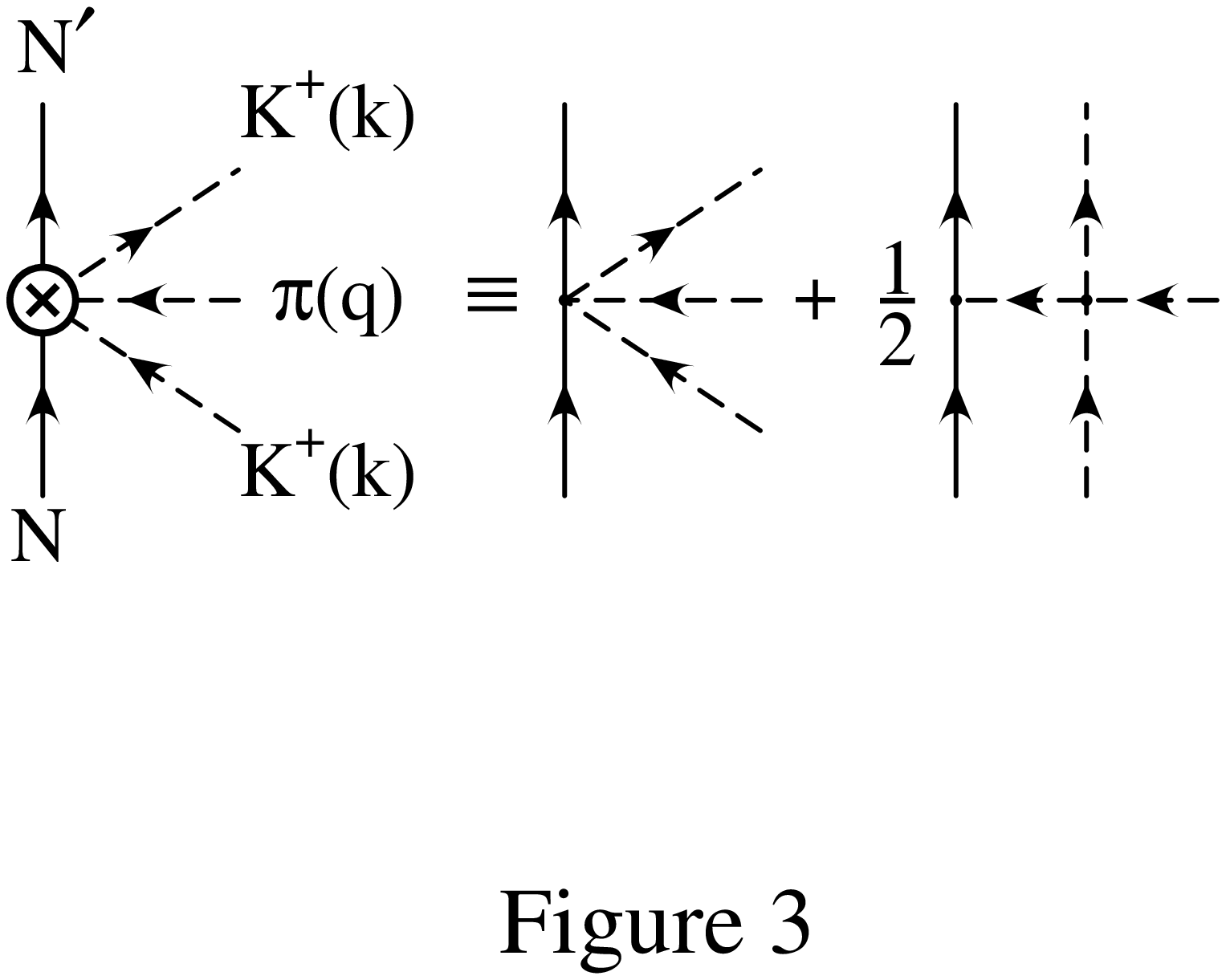}
}
\end{figure}


\newpage
\begin{figure}[bht]
\centerline{
\epsfxsize=6in
\epsffile{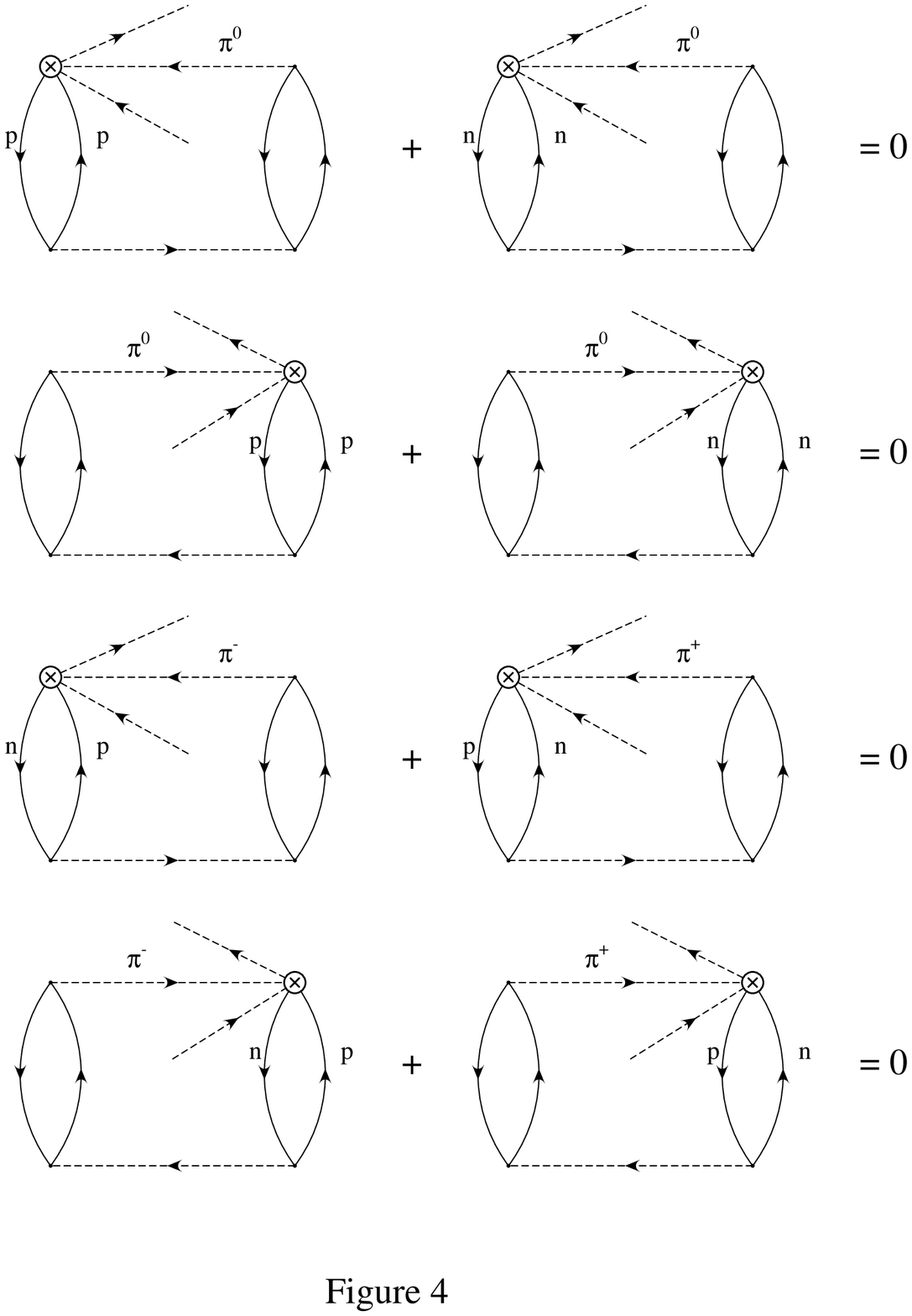}
}
\end{figure}



\bigskip

\bigskip

\begin{thebibliography}{99}
\bibitem{1} P. B. Siegel, W. B. Kaufmann and W. R. Gibbs, Phys. Rev.
C31 (1985) 2184.
\bibitem{2} C. M. Chen and D. J. Ernst, Phys. Rev. C45 (1992) 2019
\bibitem{3} S. V. Akulinichev, Phys. Rev. Lett. 68 (1992) 290
\bibitem{4} M. F. Jiang and D. S. Koltun, Phys. Rev. C46 (1992) 2462
\bibitem{5} C. Garc\'{\i}a-Recio, J. Nieves and E. Oset,
Phys. Rev. C 51 (1995) 237
\bibitem{6} G. E. Brown, C. B. Dover, P.  B. Siegel and W. Weise, Phys. Rev.
Lett. 60 (1988) 2723
\bibitem{7} C. Garc\'{\i}a-Recio, E. Oset and L. L. Salcedo, Phys. Rev.
C37 (1988) 194
\bibitem{8} R. Rockmore, Phys. Rev. C40 (1989) R13
\bibitem{9} E. Oset, C. Garc\'{\i}a-Recio and J. Nieves, Nucl. Phys. A584
  (1995) 653

\bibitem{10} J. Gasser and H. Leutwyler, Nucl. Phys. B250 (1985) 465
\bibitem{11} Ulf-G. Mei{\ss}ner, Rep. Prog. Phys. 56 (1993) 903; V. Bernard,
N. Kaiser and Ulf-G. Mei{\ss}ner, Int. J. Mod. Phys. E, in print
[hep-ph/9501384]
\bibitem{12} A. Pich, Rep. Prog. Phys. in print [hep-ph/9502366]
\bibitem{13} G. Ecker, Prog. Part. Nucl. Phys. 35, in print [hep-ph/9501357]
\bibitem{14} D. V. Bugg et al., Phys. Rev. 168 (1968) 1466; Y. Mardor et al.,
Phys. Rev. Lett. 65 (1990) 2110; R. A. Krauss et al., Phys. Rev. C46 (1992)
2019
\bibitem{15} E. Jenkins and A.V. Manohar, Phys. Lett. B259 (1991) 353
\bibitem{16} C. H. Lee, H. Jung, D. P. Min and M. Rho, Phys. Lett.
B326 (1994) 14
\end{thebibliography}
\end{document}